# Observing Teachers' Instrumental Pedagogical Orchestration in Synchronous Online Learning: A Multimodal Grid Based on Videoconferencing Traces

Intissar BAMOU[a], Christine MICHEL[b] and Hassina El KECHAI[2][c]
*University of Poitiers, TECHNE Laboratory (UR 20297), Poitiers, France*
*intissar.bamou@univ-poitiers.fr*, *christine.michel@univ-poitiers.fr* , *hassina.el.kechai@univ-poitiers.fr*

Abstract: Synchronous online teaching environments pose specific challenges for the analysis of pedagogical activity as teaching takes place via videoconferencing platforms and interactions are multimodal. While pedagogical orchestration has been extensively studied in the context of face-to-face courses and at the level of instructional design, the analysis of teaching in videoconferencing environments remains under-explored and insufficiently instrumented in terms of methodology. This article proposes a multimodal observation grid designed to analyze the instrumentalised pedagogical orchestration of teachers during synchronous online classes. Based on theories of pedagogical orchestration, multimodality and professional gestures of teachers, this grid identifies a set of observable indicators related to communicational gestures, posture, gaze and the management of digital tools. These indicators are structured and ranked in order of priority according to their observability and analytical relevance. They are operationalised to consider the constraints associated with data that can be analysed in videoconference class contexts. The proposed grid aims to provide a reproducible methodological framework for the analysis of instrumental pedagogical orchestration, with a view to future empirical validation.

## 1 INTRODUCTION

Synchronous online teaching has become a structuring modality in higher education. Its development has profoundly transformed teaching activity by reshaping interactional dynamics, regulatory processes, and the coordination of pedagogical resources. In these mediated environments, teaching becomes highly instrumented and inherently multimodal, as it unfolds through videoconferencing platforms that simultaneously mediate content delivery, interactional exchanges, and the management of digital tools.

Teachers must manage multiple actions concurrently in these environments (microphone, camera, chat, screen-sharing), resulting in a combination of verbal, bodily, and instrumental activities that characterizes the complexity of the conducting phase in videoconferencing contexts.

The concept of pedagogical orchestration provides a useful framework for understanding this complexity as a real-time coordination process involving activities, interactions, and tools (Dillenbourg & Jermann, 2010; Prieto et al., 2015). However, this notion has been widely explored in face-to-face contexts and at the level of instructional design, its enactment during the conducting phase in videoconferencing environments remains methodologically under-instrumented. A major current focus of existing protocols on primarily designed for in-person teaching, struggle to capture the multimodal and instrumental micro-actions on online learning. It therefore appears necessary to develop indicators specifically adapted to the traces produced in videoconferencing settings, to analyze the conduct of teaching as a process of multimodal and instrumented orchestration.

This paper introduces an observation framework that (1) identifies which multimodal and instrumental teacher behaviors can be effectively observed and coded from videoconferencing traces, and (2) structure these behaviors into operational indicators to construct an observation grid adapted to the technical and interactional constraints of synchronous online teaching.

[a] https://orcid.org/0009-0000-4563-5606
[b] https://orcid.org/0000-0003-3123-913X
[c] https://orcid.org/0000-0003-2025-9565

In this study, we adopt a teacher-centered methodological perspective. Previous research has shown that pedagogical orchestration is enacted through teachers in situ actions (Dillenbourg & Jermann, 2010; Prieto et al., 2015). Therefore, this study focuses on observable teacher behaviors as the main unit of analysis.

At this stage, this study does not examine the direct impact on students. Instead, the relationship between these indicators and student engagement is left for future research.

# 2 RELATED WORK

## 2.1 Instrumental Pedagogical Orchestration in Synchronous Online Learning

The concept of pedagogical orchestration allows teaching to be understood as a set of practices through which teachers organize, coordinate, and adjust learning activities in real time within a context characterized by multiple constraints (Dillenbourg, 2012). Teaching activity involves a high degree of organizational, temporal, spatial, material, and interactional complexity, requiring teachers coordinate resources, interactions, time, and tools throughout the session. Dillenbourg and Jermann (2010) defined pedagogical orchestration as a set of practices aimed at managing learning activities, distributing resources, adapting instructions, and regulating interactions. Prieto et al. (2015) broaden this definition by emphasizing that orchestration constitutes a dynamic process through which teachers design, guide, and adjust pedagogical activities according to contextual constraints.

In synchronous online learning, this orchestration is necessarily instrumental, as it unfolds through digital videoconferencing platforms. These platforms provide access to observable traces (screen sharing, microphone activation, chat use, camera display), while also imposing constraints of visibility and mediation that transform how teaching activity is enacted. Synchronous online teaching relies on multimodal resources. According to Hoffman (2018), learning situations mobilize visual, written, spoken, and bodily modalities. In videoconferencing contexts, these modalities are expressed through gaze, gestures, posture, and the use of digital (e.g., chat and screen sharing). Videoconferencing platforms such as Webex or Zoom can be considered as flexible technologies that teachers appropriate for orchestration purposes (Song, 2021). They do not prescribe a specific orchestration model but instead offer a set of functionalities that teachers mobilize and combine according to situational needs.

This technological distinction highlights the temporal dimension of orchestration, which unfolds across design and conducting phases, the latter corresponding to real-time classroom management (Tchounikine, 2013). However, most research on pedagogical orchestration focuses either on face-to-face teaching contexts or on the design phase of instruction (Song, 2021), leaving the conducting phase in synchronous online learning relatively under-explored. This gap is particularly problematic in environments where teachers must constantly adjust their actions based on partial and mediated signals.

## 2.2 Multimodality and Nonverbal Communication in Videoconferencing

Synchronous online teaching relies on a fundamentally multimodal logic. According to Hoffman (2018), learning situations simultaneously mobilize visual elements, written language, spoken language, and bodily behavior. In videoconferencing contexts, these modalities are expressed through gaze, gestures visible within the webcam frame, posture, and the use of digital tools such as chat or screen sharing.

Research on co-verbal gestures (McNeill, 1992; Tellier, 2008) and on teachers' professional gestures (Bucheton & Soulé, 2009; Duvillard, 2014) demonstrates that these resources are not peripheral; they actively contribute to structuring discourse, directing attention, and regulating interactions. They are not mere accompaniments to verbal communication but play a role in meaning making and in maintaining a shared interactional frame. Within multimodal analysis, several studies have highlighted the central role of gestures, gaze, and posture in pedagogical interactions, including in videoconferencing contexts (Cohen & Guichon, 2016; Cappellini et al., 2023; Lim, 2019).

Research specifically addressing videoconferencing, particularly in language education (Cappellini et al., 2023), is often located within scripted telecollaborative contexts and prioritizes fine-grained qualitative analyses rather than the formalization of operational indicators applicable to the study of the conducting phase in synchronous online teaching. Moreover, in videoconferencing environments, bodily manifestations are only partially visible and are shaped by webcam framing and technical constraints. Some micro-gestures become more salient, while others disappear outside the frame, making observation and interpretation more complex. An analysis of instrumental pedagogical orchestration must therefore account both for the resources mobilized by teachers and for the conditions under which these resources become visible.

## 2.3 Research Problem and Research Questions

**Limits of Existing Observation Models and Protocols.** Several protocols and tools have been developed to analyze teaching practices, including CLASS, TDOP and COPUS. These instruments make it possible to describe certain dimensions of pedagogical interaction; however, they were primarily designed for face-to-face teaching contexts and rely on relatively broad analytical categories.

In synchronous online learning, these protocols struggle to capture the fine-grained articulation between platform-mediated instrumental actions and partially visible multimodal behaviors. They document only marginally the micro-adjustments related to screen sharing management, chat use, camera regulation, or gaze distribution within the digital space of videoconferencing. There is therefore a need to develop indicators specifically adapted to the traces produced in videoconferencing environments to analyze the conducting phase as a process of multimodal and instrumental orchestration. This leads to the following question of how to formalize an operational analytical framework capable of objectifying instrumental pedagogical orchestration in videoconferencing contexts based on available multimodal traces. Considering the existing literature, instrumental pedagogical orchestration in synchronous online learning emerges as an activity that is both highly instrumental and fundamentally multimodal. Although it is widely recognized as a key lever for learning, its real-time implementation in videoconferencing remains methodologically under-instrumented.

**Research Questions.** Two research questions guide this study: RQ1. Which multimodal and instrumental teacher behaviors can be effectively observed and coded from videoconferencing traces? RQ2. How can these behaviors be structured into operational indicators to construct an observation grid adapted to the technical and interactional constraints of synchronous online learning? To address these questions, this article presents a multimodal observation grid focused on the conducting phase of teaching. By identifying and structuring observable indicators related to gestures, posture, gaze, and the technology-mediated actions, the study aims to provide an operational methodological framework for analysing teaching activity in synchronous online environments.

# 3 METHOD

## 3.1 Research Design and Methodological Positioning

This study adopts a multimodal methodological approach aimed at analysing pedagogical orchestration based on observable traces in videoconferencing contexts. The methodological objective is to operationalize instrumental pedagogical orchestration during the conducting phase of teaching through multimodal indicators observable in videoconferencing. The orchestration is understood as a dynamic and mediated process that manifests itself through the actions produced by the teacher in situ (Dillenbourg & Jermann, 2010; Prieto et al., 2015). The proposed framework does not aim to infer teachers' intentions or internal cognitive processes. Rather, it focuses on observable manifestations of coordination between verbal discourse, gestures, posture, gaze, and the management of digital tools. Instrumental pedagogical orchestration is thus approached as an observable and analysable activity, mediated by technological artefacts and enacted in real time.

## 3.2 Construction of Indicators: Conceptual Structuring

The first methodological step consisted in identifying and structuring a set of indicators capable of describing teachers' instrumental pedagogical orchestration in synchronous online learning. The construction of the indicators draws upon four complementary strands of research: studies on pedagogical and instrumental orchestration (Dillenbourg & Jermann, 2010; Tchounikine, 2013; Prieto et al., 2015); research on co-verbal and communicative gestures (McNeill, 1992; Tellier, 2013); studies on teachers' professional gestures and classroom regulation (Bucheton & Soulé, 2009); and multimodal approaches to mediated teaching and videoconferencing (Cohen & Guichon, 2016; Cappellini et al., 2023). Together, these frameworks converge toward a conception of teaching as a multimodal activity distributed between the teacher's body and the digital instruments mobilized during the session.

### 3.2.1 Observed Dimensions

The indicators are organized into four analytical dimensions corresponding to complementary registers of teaching activity in videoconferencing contexts (see Table 1).

**Communicative Gestures (I1–I5).** Communicative gestures include body manifestations that accompany or complement verbal discourse. They comprise: Iconic gestures (I1) representing a shape, size, or action; deictic gestures (I2), cursor movement, highlighting, digital pointing; rhythmic gestures (I3), structuring discourse; metaphoric gestures (I4), making abstract concepts visible; nonverbal feedback (I5), such as nodding or smiling, contributing to interactional regulation. These gestures support discourse organization, attention orientation, and socio-affective regulation within an environment where bodily cues are partially mediated.

**Posture (I6–I8).** Posture refers to component of embodied engagement and activity regulation. Three indicators are distinguished: body orientation (I6), including leaning toward the screen; bodily stability (I7), defined as maintaining an immobile position over a measurable duration; bodily mobility (I8), reflected in visible changes in position. These indicators capture the teacher's embodied dynamics within a digitally framed space constrained by webcam positioning.

**Gaze (I9–I14).** Gaze constitutes a central operator of attentional and interactional coordination in videoconferencing environments. The indicators distinguish: gaze toward the camera (I9), associated with direct address to students; gaze toward the screen or shared material (I10); alternation of gaze toward student thumbnails (I11); focused gaze on an identified student (I12); off-screen gaze (I13); gaze fixation (I14), measurable through eye-tracking data. These indicators enable analysis of the distribution of teachers' visual attention and their management of mediated interactions.

**Instrumental Gestures (I15–I19).** Instrumental gestures refer to gestures related to the management of digital tools (Cappellini et al., 2023; Cohen & Guichon, 2016). According to Cadoz (1994), an instrumental gesture constitutes a mode of communication specific to the gestural channel, distinct from the "bare" gesture, insofar as it is performed through physical interaction with a material object (the instrument). They include platform control, screen sharing, manipulation of microphone and camera, chat use and the organization of interactive activities. These indicators make observable the instrumental dimension of orchestration and how teachers manage and configure the digital environment.

Table 1 Categories, Subcategories, and Indicators of Instrumental Pedagogical Orchestration

| Category | Subcategory | Indicator | Priority |
|---|---|---|---|
| Communicative Gestures | Iconic gestures | I1 | ++ |
| | Deictic gestures | I2 | +++ |
| | Rhythmic gestures (beats) | I3 | + |
| | Metaphoric gestures | I4 | + |
| | Nonverbal feedback | I5 | +++ |
| Posture | Body orientation | I6 | ++ |
| | Bodily stability | I7 | +++ |
| | Bodily mobility | I8 | +++ |
| Gaze | Gaze toward camera | I9 | +++ |
| | Gaze toward screen | I10 | +++ |
| | Alternating gaze | I11 | +++ |
| | Focused gaze | I12 | +++ |
| | Off-screen gaze | I13 | ++ |
| | Gaze fixation | I14 | +++ |
| Instrumental Gestures | Platform manipulation | I15 | +++ |
| | Management of shared materials | I16 | +++ |
| | Written interaction (chat) | I17 | +++ |
| | Launching activities | I18 | +++ |
| | Step-by-step demonstration | I19 | +++ |

Table 1 does not constitute an exhaustive taxonomy of teaching activity, it presents a structured set of analytical categories adapted to the specific constraints of videoconferencing, based on a compromise between theoretical grounding and empirical feasibility.

### 3.2.2 Functional Value of the Indicators

Beyond their descriptive definition, the selected indicators were conceived as functional indices of instrumental pedagogical orchestration. Each indicator is thus associated with a dominant pedagogical function, linking observable teachers behaviors to the purposes of conducting a lesson in synchronous online learning. This functional approach is grounded in a conception of pedagogical orchestration as a real-time regulatory process, during which the teacher simultaneously coordinates content explanation, attention management, interaction regulation, and the management of the digital environment.

Table 2 synthesizes these analytical functions by distinguishing, for each group of indicators, the associated dominant pedagogical function and the primary mode of action mobilized (bodily, visual, or instrumental). Indicators I1 to I4 correspond to communicative gestures that directly support *explanation and conceptualization*. Iconic, rhythmic, and metaphoric gestures contribute to *structuring discourse and making abstract content accessible*, while digital deictic gestures (I2) play a central role in *directing attention toward shared materials*. These gestures embody a close coordination between bodily resources and the digital environment. Indicators I5 and I6–I8 relate to *interactional regulation and embodied engagement*. Nonverbal feedback (I5) contributes to maintaining an interactional climate and encouraging participation, while variations in posture and mobility (I6–I8) reflect adjustments in teacher engagement in response to the constraints of the videoconferencing setting. Gaze-related indicators (I9–I14) constitute a coherent set dedicated to *attention management*. Gaze orientation and fixation make it possible to analyze how the teacher distributes attention between the camera, shared materials, student thumbnails, and off-screen spaces, thereby revealing strategies of social presence, monitoring of student activity, and coordination with digital tools. Finally, instrumental indicators (I15–I19) relate to *the management of the digital environment*. They make visible the actions through which the teacher configures the videoconferencing platform, organizes materials, initiates interactive activities, and guides students in their use of digital tools. These indicators materialize the instrumental dimension of pedagogical orchestration at the core of conducting online lessons.

Table 2 Synthetic Operational Definition

| Indicator | Dominant Pedagogical Function | Mode |
|---|---|---|
| I1 | Support for explanation and making content visible | Bodily |
| I2 | Orientation of attention toward shared materials | Bodily and digital |
| I3 | Structuring of discourse | Bodily |
| I4 | Support for conceptualization | Bodily |
| I5 | Interactional regulation | Bodily |
| I6–I8 | Regulation of embodied engagement and rhythm | Bodily |
| I9–I14 | Management and distribution of attention | Visual |
| I15–I19 | Management of the digital environment (content, chat, tools) | Instrumental |

This functional reading of the indicators moves beyond a purely descriptive account of observable behaviors and orients the analysis toward the identification of orchestration configurations, understood as recurring combinations of pedagogical functions enacted in situ.

## 3.3 Operationalization of the Indicators

The second methodological step consisted in transforming the conceptual indicators into operational indicators that are observable and codable based on the traces available in synchronous online learning environments.

### 3.3.1 Operationalization

Four criteria guided this operationalization process: empirical observability, intersubjective codability, contribution to the orchestration, technical feasibility.

First, *empirical observability* required that each indicator be detectable in available data sources, this meant that the indicator had to be visible in videoconferencing recordings (Webex) or an external camera and, for gaze-related dimensions, identifiable through eye-tracking data. Indicators that could not be reliably perceived in these traces were excluded from the grid. Second, *intersubjective codability* ensured methodological robustness. Each indicator had to be defined in such a way that multiple coders could annotate it consistently and reproducibly. This criterion aimed to reduce interpretative ambiguity and to support future reliability testing, such as inter-coder agreement measures. Third, the *contribution to the orchestration* required that the indicator meaningfully inform the analysis of pedagogical coordination, showing how teachers articulate discourse, gestures, posture, gaze, and digital tools use in real time, rather than describing isolated behaviors. Finally, *technical feasibility* refers to the compatibility of each indicator with the constraints of the recording setup (e.g., camera framing, partial visibility, and variable video quality) only those indicators that could be reasonably and consistently captured within these constraints were retained.

These criteria aim to ensure a reasoned balance between analytical richness and methodological robustness.

### 3.3.2 Prioritization Criteria

Not all indicators exhibit the same degree of stability or descriptive value. A prioritization was therefore established by crossing two criteria: their effective observability within videoconferencing environments and their contribution to the description of the instrumental pedagogical orchestration. Three levels of priority were defined. *High priority (+++)* indicators are central to the analysis, as they are both clearly observable in videoconference recordings and strongly connected to the core processes of pedagogical coordination in real time. *Medium priority (++)* indicators remain analytically relevant but are more sensitive to technical constraints such as camera framing, image quality, or partial visibility, which may affect their consistency. *Low priority (+)* indicators play a complementary role; their observability is unstable, or their analytical contribution overlaps partially with more robust indicators.

The attribution of priority levels is based on the articulation between observability and contribution to the description of the instrumental orchestration. Table 1 synthesizes this operationalization by associating each indicator with its level of priority. For example, deictic gestures (I2) were classified as a high priority (+++) because they are directly visible on the screen, codable, and play a central role in coordinating discourse and shared materials. In contrast, rhythmic gestures (I3), which are more dependent on camera framing and sometimes only partially visible, were classified as low priority (+) due to unstable observability despite their role in structuring discourse. For the medium priority level (++), indicators I1, I6, and I13 were positioned at an intermediate level, as they are generally observable in videoconferencing contexts while remaining dependent on framing and technical conditions. Their contribution to the analysis of the orchestration is significant, although their observational stability remains conditioned by the technical setup.

## 3.4 Coding Procedure and Data Analysis

The following section outlines the general methodological orientation that will guide the coding procedure and data analysis. At this stage, the protocol should be understood as a structured analytical framework that remains to be empirically stabilized and validated. Several aspects—particularly segmentation criteria, annotation granularity, and reliability procedures—will require further refinement through pilot analyses.

**Corpus and context.** The analysable data will originate from synchronous online teaching sessions conducted on the Webex platform. The dataset may include video and audio recordings of the videoconferencing session. Complementary video recordings, captured via an external camera, may also be integrated to extend the observable field beyond the webcam frame and to allow a more detailed analysis of posture and gestures that might otherwise remain partially invisible. Eye-tracking data will be used to analyze the orientation and distribution of visual attention, complementing video traces and refining the analysis of gaze behavior (Duchowski, 2007).

The dataset may also incorporate platform-generated traces (e.g., chat messages, connection logs, presence indicators), which would serve as contextual information to support annotation and interpretation. Data collection will be conducted in accordance with ethical research principles, including participant information, informed consent, and appropriate data protection procedures.

**Data Segmentation.** Teaching sessions will be segmented into episodes. These episodes will be provisionally defined based on pedagogical and instrumental events, such as screen sharing, demonstrations, chat interactions, or the initiation of interactive activities. Each episode is expected to constitute a coherent unit of analysis for the annotation of multimodal indicators. However, the operational definition of episode boundaries will need to be empirically tested and possibly adjusted to ensure consistency and analytical relevance.

**Multimodal Annotation.** Data will be manually annotated using ELAN, with multiple annotation tiers corresponding to gestures, gaze, posture, and instrumental actions. Depending on the nature of each indicator, annotation may take the form of discrete events (e.g., pointing gestures, microphone activation) or temporal intervals (e.g., posture orientation, gaze fixation). The annotation scheme will require iterative refinement through pilot coding sessions. The annotation scheme will be refined through pilot coding to improve consistency and reliability.

**Methods of Analysis.** The analysis combines two complementary levels. A quantitative component will focus on frequencies, duration, and co-occurrences of indicators, to identify recurrent patterns. A qualitative component will concentrate on the interpretation of multimodal configurations and sequences of pedagogical orchestration within specific episodes.

Together, they provide both a global and a fine-grained understanding of pedagogical orchestration in instrumented environment.

This analytical framework provides a structured basis for the empirical validation of the proposed grid,

particularly for pattern detection, temporal modelling, and cross-session comparison.

**Empirical validation protocol.** The empirical validation will rely on the ability of the observation grid to generate the measures in a systematic and reproducible manner, as well as to identify stable and interpretable configurations that can shed light on the identification and understanding of teachers' multimodal and instrumental behaviors in videoconferencing contexts. To assess the *reliability of the coding scheme*, a double-coding procedure will be conducted on a subset of the corpus by two independent annotators using the same annotation protocol. Inter-coder agreement will be quantified using Cohen's Kappa. Agreement will be calculated on time-aligned annotation units, and Kappa values will be reported for each category. A threshold of $\kappa \geq 0.60$ will be considered acceptable. *Content validity* will be assessed by submitting the observation grid to a panel of experts in didactics, educational technologies, activity analysis, and multimodality, to evaluate the clarity, relevance, exhaustiveness, and non-redundancy of the categories; feedback will be analyzed based on the level of agreement among experts and the proposed adjustments. *Ecological validity* will be examined through self-confrontation interviews with teachers, based on annotated sequences, to verify the correspondence between the identified behaviors, the activity as actually experienced, and the teachers' intentions, as well as to identify any missing dimensions. Finally, the *analytical usefulness* of the method will be assessed by engaging researchers to evaluate the extent to which the results produced (frequencies, durations, co-occurrences, configurations) can be interpreted, support the development of typologies, and contribute to understanding forms of teachers' orchestration, notably by considering the degree of convergence across analyses.

# 4 SCOPE AND LIMITATIONS

The proposed methodology provides a structured way analyze teachers' instrumental pedagogical orchestration in videoconferencing environments through observable multimodal behaviors. It focuses on the analysis of multimodal coordination as it manifests on screen, considering the specific technical and interactional constraints of synchronous online teaching. The observation grid constitutes an intermediate methodological instrument, which enables systematic coding and supports both quantitative and qualitative analyses.

The grid captures observable actions but not teachers' intentions or cognitive processes, and remains dependent on technical visibility, as behaviors outside the camera frame cannot be observed. In addition, the relationship between orchestration indicators and student engagement is not addressed at this stage and requires complementary empirical investigation.

# 5 CONCLUSIONS AND FUTURE WORK

This article contributes to the study of instrumental pedagogical orchestration in synchronous online learning by proposing a methodological framework specifically adapted to videoconferencing environments. Grounded in theories of pedagogical orchestration, multimodality, and teachers' professional gestures, it introduces a multimodal observation grid focused on the conducting phase of teaching. The primary contribution lies in the identification and structuring of observable indicators derived from videoconferencing traces. By integrating communicative gestures, posture, gaze, and instrumental actions, the framework enables a systematic analysis of how teachers coordinate discourse, embodied resources, and digital tools in real time. The functional articulation of indicators—linked to explanation support, attention management, interactional regulation, and digital environment management—offers an intermediate analytical level between micro-interactional description and operational research design.

Future work will first need to focus on methodological consolidation. The coding procedure and data analysis protocol require further stabilization through pilot studies. A temporal analysis could support modelling orchestration dynamics, including with eye-tracking data. Another direction concerns the relationship between teacher orchestration and student engagement. Once stabilized, the grid may be applied to comparative studies involving different teacher profiles, disciplinary contexts, or instructional formats. Such applications would allow the identification of recurring orchestration configurations and support a transition from isolated indicators to patterned models of pedagogical coordination. A dynamic, temporally oriented analysis of sessions could further contribute to modelling the evolution of orchestration across a lesson, including phases of instrumental intensification or shifts in attentional focus, particularly when combined with eye-tracking data. Another important direction concerns the articulation between teacher orchestration and student engagement. By relating orchestration indicators to student behaviors, future research may explore the relationships between coordination strategies and engagement dynamics, moving toward relational and explanatory modelling. In the longer term, these developments could contribute to the construction of a comprehensive model of instrumental pedagogical orchestration, potentially supporting predictive analyses and informing evidence-based recommendations for teacher training in synchronous online environments.

## ACKNOWLEDGEMENTS

This work was carried out as part of TALISMAN project (ANR-22-CE38-0007).